\documentclass[twocolumn,preprintnumbers, amsmath, amssymb]{revtex4}

\def\be{\begin{equation}}
\def\ee{\end{equation}}
\def\beq{\begin{eqnarray}}
\def\eeq{\end{eqnarray}}

\def\bay{\begin{array}}
\def\eay{\end{array}}

\begin{document}
\title{The Significance of the General Principle of
Relativity}

\author{Sanjay M. Wagh}
\affiliation{Central India Research Institute, \\ Post Box 606,
Laxminagar, Nagpur 440 022, India\\
E-mail:cirinag\underline{\phantom{n}}ngp@sancharnet.in}

\date{February 17, 2005}
\begin{abstract}
In this note, we discuss the significance of the general principle
of relativity for a physical theory that abandons the newtonian
concept of force and, hence, uses an entirely different conception
for the ``cause'' behind motions of material bodies.
\end{abstract}

\maketitle \newpage
\section{Conceptual Preliminaries} \label{idea-basis}

To begin with, let us note that the foremost of the concepts
behind Newton's theory is, undoubtedly, (Galileo's) concept of the
inertia of a material body. We postulate that every material body
has this inertia for motion.

The association of the inertia of a material body with the points
of the Euclidean space is the first primary physical conception
that is necessary for Newton's theory to describe motions of
physical bodies. Then, with this association, the Euclidean
distance becomes the {\em physical distance\/} separating material
bodies and the Euclidean space becomes the {\em physical space\/}
for any further considerations of Newton's theoretical scheme.

Next, a {\em physical clock\/} is a material body undergoing
periodic motion or a periodic phenomenon. Essentially, in Newton's
theory, a physical clock is a set of points of the Euclidean space
exhibiting periodic motion under a {\em periodic\/}
transformation. Mathematically, in Newton's theory, let $A$ be the
set of all points $x_{_A}$ of the Euclidean space making up the
clock. Let $T$ be the {\em periodic\/} transformation such that
$\forall \;x_{_A}\in A,\,T^n\,x_{_A}=x_{_A}$, where $n$ is the
period of the transformation $T$.

In Newton's theory, an observer can {\em observe\/} the {\em
entire periodic motion\/} of the material body of the clock (under
the use of the transformation $T$) without disturbing the clock in
any manner whatsoever. Then, the {\em known\/} state or the
reading of the clock {\em represents\/} the {\em physical time}.
Any ``measurement'' of the physical time gives the period or the
part of the period $n$ of the transformation $T$. It is tacitly
assumed in these considerations that the involved quantities are
{\em exactly\/} measurable.

Now, consider a material point with an initial location
$\vec{x}_o$. In Newton's theory, the trajectory of this material
point is a (continuous) sequence of points of the Euclidean space.
It is then a ``curve'' traced by the point $\vec{x}_o$ under some
transformation $\tilde{T}_t$ of the Euclidean space where
parameter $t$ labels points of the sequence. Of course, the
transformation $\tilde{T}_t$ need not be periodic.

The label parameter $t$ can then be made to correspond to the
physical time in a one-one correspondence. This is theoretically
permissible as the measurement of the physical time does not
disturb the clock in any manner whatsoever in Newton's theory.
This correspondence is the physical meaning of the labelling
parameter $t$.

An observer can thus check the position of another material point
against the state of a {\em physical clock}. Without this
``correspondence,'' the geometric curve of the Euclidean space has
no physical sense for the path of a material point.

When such physical associations are carried out, we say that the
material point is at ``this'' location given by the three space
coordinates and the physical clock is {\em simultaneously\/}
showing ``this'' time. This {\em simultaneity\/} is inherent in
the physical associations of Newton's theory. It is an important
assumption of Newton's theory.

Also, as a consequence of the fact that Newton's theory treats
material bodies as {\em existing\/} independently of the space,
the state of a physical clock or its reading is {\em assumed\/} to
be independent of the motion of another material point or points
separately under considerations. Then, physical time is {\em
independent\/} also of the coordination of the metrically-flat
Euclidean space.

But, then, the motion of a material point in the ``physical''
space does not produce any change in that space. Clearly, this
fact applies also to the periodic motion or the periodic
phenomenon making up the physical clock.

The ``physical'' construction of the coordinate axes and clocks
must also be using the material bodies, for example, coordinate
axes could be constructed using ``sufficiently long'' material
rods, say, of wood. Then, any material object, a road-roller, say,
crossing the coordinate axis must ``affect'' the corresponding
wooden rod.

But, in Newton's theory, the coordinate axes of the Euclidean
space do not get affected by the motions of other material bodies.
Clearly, use of non-cartesian coordinates does not change this
state of affairs with Newton's theory.

Now, any difference of coordinates in the Euclidean space is a
``measuring stick or rod'' that can be used to ``measure'' the
physical separation of material bodies.

Furthermore, in Newton's theory, each observer has a coordinate
system of such measuring rods and clocks. Then, when one observer
is in motion (relative to another one), the {\em entire\/} system
of coordinate axes and clocks is also carried with that observer
in motion.

It needs to be adequately recognized that it is really the {\em
theoretical possibility\/} of making a {\em genuine\/} physical
measurement at any stage which is the issue involved herein.

Conceptually, the aforementioned physical situation is
``acceptable'' except in one case. Surely, we cannot have a
material rod with one observer and {\em simultaneously}, another
material rod, {\em at the same place}, with other observer in
(uniform or not) motion relative to the first one.

But, in Newton's theory, measuring stick of one observer does not
collide with that of another observer in motion even when both
these sticks arrive at the same place. Unacceptably, any two such
measuring sticks just pass through each other without even
colliding on their first contact.

The same situation does not arise for other material bodies which
are supposed to collide on their first contact. Then, in Newton's
theory, the measuring rods of the physical space - Euclidean space
- are treated {\em separately\/} than other material objects. But,
measuring rods must also be made up of material objects. Then,
their separate treatment is, theoretically, not appropriate one.
Surely, this problem with Newton's theory is, undeniably, of
serious theoretical concern.

(This above issue would not have been relevant if the Euclidean
distance were also not, simultaneously, the physical distance
separating material bodies. Mathematically, the continuum
$\mathbb{R}\times\mathbb{R}\times \mathbb{R}$ can be assumed.)

Hence, Newton's theory attempts to {\em explain all phenomena as
relations between objects existing in Euclidean space and time}.
It achieves this mainly by attributing {\em absolute\/} properties
to the space and the time, thereby totally separating them from
the properties and motions of matter.

Thus, limitations of Newton's theoretical scheme (providing his
famous three laws of motion) originate in its use of the Cartesian
concepts related to the Euclidean space and the associations of
properties of material bodies with the points of this
metrically-flat space.

Now, the entire physical structure of Newton's theory is woven
around only two basic concepts, namely, those of the inertia and
the force.

Clearly, the force, as a cause of motion, is another pivotal
concept of Newton's theory. Hence, consider the status of the
concept of force, the cause behind motions of material bodies,
within Newton's theory of mechanics.

Firstly, we could ask: What is the {\em cause\/} of this force?
Within Newton's overall theoretical scheme, only a material point
can be the source of force. A material point ``here'' {\em acts\/}
on a material point located ``there'' with the specified force.
Newton's theoretical scheme is therefore an {\em action at a
distance\/} framework.

Then, in Newton's theory, we can consider a physical body as one
material point and also other physical bodies as other material
points. We vectorially add the forces exerted by each one on the
first physical body to obtain the total force acting on it. It is
this {\em total force\/} that is used by Newton's second law of
motion to provide the means of establishing the path followed by
that physical body under the action of that total force.

In Newton's second law of motion, we {\em must\/} first {\em
specify\/} the {\em force\/} acting on a material body. Only then
can we solve the corresponding differential equation(s) and
obtain, subject to the given initial data, the path of the
material point representing that material body.

Then, without the {\em Law of Force}, it is clear that the {\em
Law of Motion\/} is {\em empty\/} of contents in Newton's
theoretical framework. This is an extremely important issue for a
physical theory.

From our ordinary, day-to-day, observations, we notice that
various objects fall to the earth when left ``free'' in the air.
We then say that objects {\em gravitate\/} towards the Earth. This
is, in a nutshell, the {\em phenomenon of gravitation}.

We then need to explain as to why the objects ``ordinarily''
gravitate to Earth, {\em ie}, why they have a tendency to come
together or why the distance between them decreases with time.

In Newton's theory, only the {\em force\/} ``causes'' the motions
of material bodies. Then, the gravitating behavior of objects is
``explainable'' only by {\em postulating\/} a suitable {\em force
of gravity\/} that makes material bodies fall to the Earth.

But, in Newton's theory, a force acts between {\em any two\/}
separate material points possessing the {\em required source
property\/} by virtue of which the force in question is generated.
Furthermore, for the internal consistency of Newton's scheme, the
force so generated by one material point on the second material
point must also be equal in amplitude but opposite in direction to
that generated by the second material point on the first material
point. This is Newton's third law of motion. This law also has the
status of a postulate within the overall scheme of the newtonian
mechanics.

Newton had {\em assumed\/} that the force of gravity is
proportional to the {\em inertias\/} of the two material points
under consideration because, following Galileo, he had postulated
that inertia ``characterizes'' a physical body.

Such a force of gravity must then be generated by a chosen body
(Earth) on {\em all\/} the other material bodies because, by
postulate, every material body possessed inertia. The force of
gravity must then be {\em universal\/} in character.

To explain many of the day-to-day observations involving the
terrestrial bodies as well as planetary motions that were already
known in details, Newton was therefore compelled to state a Law of
Force - Newton's Law of Gravity - to explain the phenomenon of
gravitation.

In fact, in Newton's theory, a material body has two {\em
independent\/} attributes: the first, its {\em inertial mass}, is
a measure of the opposition it offers to a change in its state of
motion, and the second, its {\em gravitational mass}, is a measure
of the property by virtue of which it produces the force of
gravity on another material point.

Various observations, since Galileo's times, then indicate
\cite{0411052} that the inertial and the gravitational masses of a
material body are equal to a high degree of accuracy. However,
this equality becomes an assumption of Newton's theory.

Furthermore, the inverse-square dependence of the gravitational
force on the distance separating two bodies is also an important
assumption of Newton's theory.

In relation to the inverse-square dependence of Newton's force of
gravity on distance separating two bodies, we could then always
raise questions: Why not any other power of distance? Why should
this force not contain time-derivatives of the space coordinates?
Clearly, Newton's theory offers no explanation for even the
inverse-square dependence of the force of gravitation.

Hence, in Newton's scheme, his law of gravitation has the status
of a {\em postulate\/} about the {\em force\/} acting between two
material particles separated by some spatial distance.

Also, Coulomb's law from the electrostatics provides another,
postulated, fundamental force. It is also assumed to exist {\em
universally\/} between any two {\em charged\/} material bodies. It
is an ``additional'' force, over and above that of gravity, which
Newton's theory postulates to explain the motions of charged
material points.

Now, every object does not fall to the Earth. So, ``something''
opposes the attractive force of gravity. That ``something'' must
also be another force. As an example, the force of electrostatic
repulsion can balance the force of gravity between two charged
material bodies.

Then, under the action of this ``another force'' the distance
separating two material bodies can also decrease. As an example,
the force of electrostatic attraction between two charged material
bodies can cause the decrement in their separating distance.
Therefore, not every decrement in distance between material bodies
is due to the force of mutual gravitation.

Thus, in a nutshell, one force can oppose another force. But, {\em
every\/} force must be an assumption of Newton's theory.

Then, if we find that some physical body, for example, a star, is
stable, we could, in Newton's theory, explain its stability by
postulating another suitable {\em force \/} which counterchecks
the force of self-gravity of the star. On the other hand, if the
star were unstable, existence of ``unbalanced'' forces in the star
is implied.

In Newton's theory, there are no fundamentally important issues
involved here than those related to finding the nature of the
force opposing the self-gravity of the star. It is essential to
recognize this important aspect of Newton's theory.

Nonetheless, in spite of it being an assumption of Newton's
theory, Newton's inverse-square law of gravitation does possess
certain experimental justification - it is this inverse-square
dependence that is known to be consistent with various
observations and experiments.

Still, it cannot be denied that Newton's law of gravitation is an
important {\em assumption\/} of the newtonian mechanics.

To reemphasize the status of the laws of the force in Newton's
theory, we note that every force is an assumption of this theory.
Some forces are assumed to exist {\em universally\/} between any
two material bodies. In particular, ``the force of gravitation''
is postulated by this theory.

In Newton's theory, every ``fundamental'' notion of the force
necessarily requires a source property to be attributed to
material bodies. Then, the action-at-a-distance force has this
important characteristic always.

Obviously, Newton's theory cannot hope to ``explain'' the origin
of any of such source attributes, each of these source attributes
being an assumption of that theory. Evidently, the same applies to
other action-at-a-distance theories. It is important to recognize
this fact at this early stage of our present considerations.

Perhaps, we would have been satisfied even with these assumptions
of Newton's theory if it were not for the fact that Newton's
theory does not explain the phenomena displayed by Light.
Moreover, various observations related to the wave-particle
duality of light as well as matter are also unexplainable within
the newtonian scheme.

Apart from various fundamental reasons of theoretical nature as
discussed earlier, it is also for such experiments or observations
which cannot be explained by Newton's theory, that some suitable
``new'' theory becomes a necessity.

(Historically, when various experimental discoveries contradicted
Newton's theory, the emphasis of the theoretical research suddenly
changed from that of working out the detailed implications of
Newton's theory to the quest for a ``new theory.'' However,
various serious theoretical difficulties of this theory were, of
course, quite well known since Newton's own times.)

Of course, different results of Newton's theory which successfully
describe motions of material bodies must be obtainable within the
new theory in some suitable way.

In view of the overwhelming experimental data unexplained by
Newton's theory, ``new theory'' is inevitable. Then, in order to
formulate a suitable new theory, we need to, not just modify but,
completely abandon some newtonian concepts at a fundamental level.

In the passing, let us also note here the {\em principle of
relativity\/} in Newton's theory. This principle states that: {\em
If a coordinate system $K$ is chosen so that Newton's laws of
motion hold good without the introduction of any pseudo-forces
with respect to this frame then, the {\em same\/} laws also hold
good in relation to other coordinate system $K'$ moving in {\em
uniform\/} translation relatively to $K$}. This principle is a
direct consequence of the experiments conducted by Galileo and
inferences that can be drawn from these experiments.

These issues then bring us to the question of the status of the
principle of relativity in a theoretical framework that abandons
the newtonian concept of the force. It is to this and other
related issues that we now turn to.

\section{The General Principle of Relativity}

To incorporate the physical description of the phenomena displayed
by Light, zero rest-mass object, Einstein modified the newtonian
principle of relativity as: {\em If a coordinate system $K$ is
chosen so that physical laws hold good in their simplest form with
respect to this frame then, the {\em same\/} laws also hold good
in relation to other coordinate system $K'$ moving in {\em
uniform\/} translation relatively to $K$}. This is the {\em
special principle of relativity}. As is well known, together with
the principle of the constancy of the speed of light in vacuo, it
leads to the special theory of relativity.

Einstein's this special principle of relativity is essentially the
{\em same\/} as the principle of relativity of Newton's theory.
The word ``special'' indicates here that the principle is
restricted to the case of uniform translational motion of $K'$
relative $K$ and does not extend to non-uniform motion of $K'$ in
relation to the system $K$.

But, even the special theory of relativity is not sufficiently
general to offer explanations for various physical phenomena as
observed. Not only gravity, but, as should be amply clear, the
origin of inertia as well as the origin of electrostatic charge
are also not explainable in special relativity.

Primarily, the special theory of relativity is an {\em
extension\/} only of the newtonian laws to incorporate the laws of
motion for material bodies with vanishing inertia \cite{100-yrs}.
It achieves this extension by acknowledging the fact that, in our
day-to-day experiences, we use Light to observe.

But, the special theory of relativity also rests on the metrically
flat continuum and is, thereby, beset with the problems of
treating the measuring rods and clocks separately from all other
objects. There is therefore the need to extend the special
principle of relativity.

Then, Einstein extended \cite{ein-dover} this principle on the
basis of Mach's reasoning as follows.

Mach's reasoning concerns the following situation. Consider two
{\em identical\/} fluid bodies so far from each other and from
other material bodies that only the {\em self-gravity\/} of each
one needs to be considered. Let the distance between them be
invariable, and in neither of them let there be ``internal
motions'' with respect to each other. Also, let either body, as
judged by an observer at rest relative to the other body, {\em
rotate\/} with constant angular velocity about the line joining
them. This is, importantly, a {\em verifiable\/} relative motion
of the two identical fluid bodies.

Now, using surveyor's instruments, let an observer at rest
relative to each body make measurements of the surface of that
body. Let the revealed surface of one body be {\em spherical\/}
and of the other body be an {\em ellipsoid of revolution}.

The question then arises of the {\em reason\/} behind this
difference in these two bodies. Of course, no answer is to be
considered satisfactory unless the given reason is {\em
observable}. This is so because the Law of Causality has the
``genuine scientific'' significance only when observable effects
ultimately appear as causes and effects.

As is well known, Newton's theory as well as the special theory of
relativity require the introduction of {\em fictitious or the
pseudo\/} forces to provide an answer to this issue. The reason
given by these two theories is, obviously, entirely unsatisfactory
since the pseudo-forces are unobservable.

Any cause within the system of these two bodies alone will not be
sufficient as it would have to refer to the absolute space only.
But, the absolute space is necessarily unobservable and,
consequently, any such ``internal'' cause will not be in
conformity with the law of causality.

The only satisfactory answer is that the cause must be outside of
this physical system, and that must be referred to the {\em real
difference\/} in motions of {\em distant\/} material bodies
relative to each fluid body under consideration.

Then, the frame of reference of one fluid body is {\em
equivalent\/} to that of the other body for a description of the
``motions'' of other bodies. As Mach had concluded, no observable
significance can be attached to the cause of the difference in
their shapes without this equivalence.

{\em The laws of physics must then be such that they apply to
systems of reference in any kind of motion (without the
introduction of any fictitious causes or forces)}. This is then
the extended or the {\em general principle of relativity}.

{\em Clearly, the reference frames must be constructed out of
material bodies and any motions of ``other'' material bodies must
affect the constructions of the reference frames. Therefore, the
general principle of relativity also means that the laws of
physics must be so general as to incorporate even these situations
in their entirety.}

Now, equally important is the fact that {\em the notion of the
{\em physical time\/} must undergo appropriate changes when the
above is implemented}. In particular, the correspondence of the
labelling parameter of the ``path'' of a physical body with the
time displayed by a physical clock must be different than that in
Newton's theory or in special relativity. {\em Notably, the
underlying continuum and the physical space are then different}.

Einstein connected the general principle of relativity with the
observation that a possible {\em uniform\/} gravitation imparts
the same acceleration to all bodies. This insight leads us to
Einstein's equivalence principle. It arises as follows.

Let $K$ be a Galilean frame of reference relative to which a
material body is moving with uniform rectilinear motion when far
removed from other material bodies. Let $K'$ be another frame of
reference which is moving relatively to $K$ in uniformly
accelerated translation. Then, relatively to $K'$, that same
material body would have an acceleration which is independent of
its material content as well as of its physical state.

The observer at rest in frame $K'$ can then raise the question of
determining whether frame $K'$ is ``really'' in an accelerated
motion. That is, whether this is the only cause for the
acceleration of bodies being independent of material content.

Now, let various bodies, of differing material contents and of
differing inertias, fall freely under the action of Earth's
gravity after being released from the same distance above the
ground and at the same instant of time. Galileo had, supposedly at
the leaning tower of Pisa, observed that these bodies reach the
ground at the same instant of time and had thereby concluded that
these bodies fall with the {\em same\/} accelerations.

Hence, the {\em decrement in distance\/} between material bodies
displaying only the phenomenon of gravitation is then {\em
uniquely\/} characterized by the fact that the acceleration
experienced by material bodies, occupying sufficiently small
region of space near another material body of large spatial
dimension, is independent of their material content and their
physical state. Here, the gravitational action of the larger
material body can then be treated as being that of {\em uniform\/}
gravitation.

Therefore, the answer to the question raised by the observer at
rest in the frame $K'$ is in the negative since there does exist
an analogous situation involving the phenomenon of uniform
gravitation in which material bodies can possess acceleration that
is independent of their material content and the physical state.

Thus, the observer at rest in the frame $K'$ can alternatively
explain the observation of the ``acceleration being independent of
the physical state or the material content of bodies'' on the
basis of the phenomenon of {\em uniform\/} gravitation.

The mechanical behavior of material bodies relative to the frame
$K'$ is then the same as that in the frame $K$, supposedly being
considered as ``special'' as per the special principle of
relativity. We can therefore say that the two frames $K$ and $K'$
are {\em equivalent\/} for the description of the facts under
consideration. Clearly, we can then extend the special principle
of relativity to incorporate the ``accelerated'' frames.

Borrowing Einstein's words on this issue \cite{ein-dover}, this
above situation is then {\em suggestive\/} that {\em the systems
$K$ and $K'$ may both with equal right be looked upon as
``stationary,'' that is to say, they have an equal title as
systems of reference for the physical description of phenomena}.
[Note the use of the word ``suggestive'' in this statement.]

Now, the equivalence of inertial and gravitational masses of a
material body refers to the ``equality'' of corresponding
qualities of a material body. But, this is permissible only in a
theory that assumes the concept of a force as an external cause of
motions of material bodies. The concept of the gravitational mass
is, however, {\em irrelevant\/} when the concept of force is
abandoned. Only the concept of the inertia of a material body is
then relevant to the motions of physical bodies.

What then is the status of the general principle of relativity in
a theory that completely abandons the concept of force? Does it
hold in the absence of the concept of force?

From the above, it should now be evident that the general
principle of relativity stands even when the concept of force is
abandoned because it only deals with the {\em observable\/}
concept of an acceleration due to gravity. It rests only on the
observation that uniform gravity imparts the same acceleration to
all the bodies. The concept of force is not at all essential to
establish this fact.

Now, it is crucial to recognize that the equivalence principle
establishes only the {\em consistency\/} of the phenomenon of
gravitation with that of the general principle of relativity.
Clearly, the equivalence principle {\em is not logically
equivalent\/} to the general principle of relativity.

As noted earlier, Einstein had, certainly, been quite careful to
use the word ``suggestive'' in stating the relation of these two
{\em different\/} principles. He further wrote in \cite{ein-dover}
``... in pursuing the general theory of relativity we shall be led
to a theory of gravitation, since we are able to ``produce'' a
gravitational field merely by changing the system of
coordinates.''

From the above discussion, it should be equally clear that the
general principle of relativity can be reached from more than one
vantage issues. Each such issue can then indicate only that some
physical phenomenon related to that issue is {\em consistent\/}
with this principle of relativity. The mutual consistency of the
general principle of relativity and various physical conceptions
then becomes the requirement of a satisfactory theory.

It must therefore be realized that the physical construction of
the frames of reference, the {\em physical coordination\/} of the
{\em physical space\/} using measuring rods, is the primary
requirement of the satisfactory theory based on the general
principle of relativity - the general theory of relativity.

Then, it should now be also clear that the {\em general theory of
relativity}, a physical theory explicitly based on the general
principle of relativity, will not be just a theory of gravitation
but, of necessity, also the {\em theory of everything}. It is
certainly decisive to recognize this.



{\em Therefore, a theory which abandons the concept of force
completely can ``explain'' the phenomenon of gravitation by
demonstrating that the decrement of distance between material
bodies is, {\em in certain situations}, independent of their
material contents and physical state.} By showing this, a theory
of the aforementioned type can incorporate the phenomenon of
gravitation.

Why is this above mentioned demonstration expected to hold only in
{\em certain situations}?

To grasp the essentials here, let us recall that, in Newton's
theory, only the {\em total force\/} acting on a physical body is
used by Newton's second law of motion. We usually also {\em
decompose\/} this force into different parts in the well known
manner as the one arising due to gravity, the one arising due to
electrostatic force etc.

We have become so accustomed to this classical newtonian thinking
in terms of the aforementioned decomposition of the total force
that we ignore the following important fact.

What matters in Newton's theory for the motion of any physical
body is the total force acting on it and not the decomposition of
this total force in parts, the decomposition, strictly speaking,
being quite {\em irrelevant}.

Thus, the phenomenon of gravitation is, then, ``displayed'' by
material bodies, essentially, only in {\em certain situations},
those for which the total force is that due to gravity.

This above is, in overall, the significance of the general
principle of relativity.

\section{General Expectations from the New Theory}

Now, what are our general expectations from any ``new'' theory
then?

Clearly, this ``new'' theory is {\em not\/} the special theory of
relativity or even the known quantum theory. As we have already
seen before, the conceptual framework of special relativity is
{\em not\/} sufficiently general. The quantum theory too has not
the required general basis as its framework needs to specify
inertia and charge.

The standard formalism of the quantum theory provides us,
essentially, only the means of calculating the probability of a
physical event involving physical object(s). It presupposes
therefore that we have specified, either the lagrangian or the
hamiltonian, {\em ie}, certain physical characteristics of the
problem under consideration. Evidently, this is necessary to
determine Schr\"{o}dinger's $\Psi$-function using which we can
then make (probabilistic) predictions regarding that physical
phenomenon under consideration.


Therefore, the formalism of the quantum theory, leading us to the
probability of the outcome of a physical experiment about a chosen
physical object, cannot provide us the means of ``specifying''
certain intrinsic properties of that physical body. This fact,
precisely, appears to be the reason as to why we have to {\em
specify by hand\/} the values of the mass and the charge in
various operators of the non-relativistic as well as relativistic
versions of the quantum theory.

The ``origins'' of ``intrinsic'' properties of physical bodies
cannot then be explainable on the basis of the quantum theory.
Then, the formalism of quantum theory as well its inherently
probabilistic considerations cannot provide the {\em universal\/}
basis for physics, in general.

Now, the concept of the inertia of a material body is {\em more
fundamental\/} than that of the force because the conception of
gravitational force requires the introduction of gravitational
mass which is conceptually very different but ``equals'' the
inertia in value to a high degree of accuracy \cite{0411052}.
Hence, only the newtonian concept of force comes under scrutiny
for modifications.

Therefore, it must be adequately recognized that the newtonian
concept of ``force'' will have to be abandoned in the process of
developing the new theoretical framework. In other words, the
``cause'' behind the motions of material bodies will have to be
conceptually entirely different than has been considered by
Newton's theory.

Consequently, ``agreements'' of the results of the new theory with
the corresponding ones of Newton's theory can only be {\em
mathematical\/} of nature. The physical conceptions behind the
mathematical statements of the new theory will not be those of
Newton's theory.

Therefore, any explanation of the phenomenon of gravitation in the
new theory will only involve the demonstration of the ``decrement
of distance'' under certain situations involving material bodies.
It must, of course, be shown that this decrement in distance is,
for these situations, such that the ``acceleration'' of the bodies
is independent of their material content and the physical state.
It must also be shown that the ``known'' inverse-square dependence
of this phenomenon arises in the new theory in some mathematical
manner.

Any such ``new theory'' must then explain the ``origin'' of the
inertia of material bodies. It must also incorporate the
``physical'' construction of the coordinate system that must,
necessarily, change with the motions of material bodies in the
``physical'' space. Without the appropriate incorporation of these
two issues, no theory can be considered to be physically
satisfactory.

Any such ``new'' theory needs also ``explain'' the equality of the
{\em inertial\/} and the {\em gravitational\/} mass of a material
body. The equality of these two entirely different physical
conceptions, even with experimental uncertainties, is a sure
indication that the {\em same quality\/} of a material body
manifests itself, according to circumstances, as its inertia or as
its weight (heaviness).

But, it must be remembered that the concept of the gravitational
mass owes its origin to the newtonian concept of the force.
Gravitational mass is the source of the newtonian gravitational
force. Also, electrostatic charge is the source of Coulomb's
electrostatic force.

But, the ``source properties'' cannot be basic to the new theory
that abandons the concept of the force. Consequently, the
gravitational mass of the material body will not be fundamental to
the new theory, but the inertial mass will be. Some ``entity''
that replaces the electrostatic charge will also be basic to the
new theory.

The equality of the gravitational mass and the inertial mass for a
material body is then not really the issue for the new theory.

But, it must also follow from the mathematical framework of the
new theory that the inertial mass can also be ``naturally''
considered as the ``source'' in the mathematical quantity that can
be the newtonian gravitational force.

Similarly, the quantity that, in the new theory, replaces the
electrostatic charge must also naturally appear as the ``source''
in the mathematical quantity that can be considered to be
Coulomb's electrostatic force.

Furthermore, we need to demand that the ``new'' theory must also
not contain the law of motion which is ``independent'' of the law
of the force. That is to say, the force as an external quantity,
to be ``specified'' separately of the law of motion, must not
occur in this new theory. In it, we can only have the law of
motion.

Crucially, abandonment of the concept of force that is independent
of the law of motion applies, at the same time, to ``every kind of
(fundamental) force'' postulated to be acting between the material
particles by Newton's theory.


Therefore, the {\em conceptual framework as well as the
mathematical formalism or procedure\/} by which we ``replace'' the
concept of force (as an ``external cause of motion'' that is
independent of the law of motion) {\em will have to be applicable
to every kind of (fundamental) force that Newton's theory has to
postulate or assume to explain the observed motions of material
bodies}.

The {\em Principle of the Simplicity\/} (of Theoretical
Construction) dictates that this above must be the case for a
satisfactory theory.

Replacing only the concept of the gravitational force is then
unacceptable not only from this point of view of the simplicity
but also because the resultant theory then cannot account for the
entirety of charged material bodies within its hybrid framework.
Charged material bodies will have to be the singularities of the
electrostatic force but not of the gravitational force in the
mathematical framework of such a hybrid theory. Any such hybrid
framework is then bound to be physically inconsistent and, hence,
unacceptable.

Then, the mathematical procedure by which we replace the notion
of, say, Newton's gravitational force {\em cannot\/} be expected
to be entirely {\em different\/} than the one adopted, say, for
replacing the notion of Coulomb's electrostatic force.

Now, the concept of force is, in a definite mathematical sense
\cite{dyn-sys}, {\em equivalent\/} to that of certain
transformations of the point of the (Euclidean) space in Newton's
theory. This is then suggestive that mathematical transformations
of points of the (underlying continuum) space can, quite generally
as well as naturally, ``replace'' the newtonian concept of force
as a cause of motion.

It should then be evident that the mathematical laws obtainable in
this way will be applicable to {\em every reference frame}, and,
hence, this mathematical formalism will be in conformity with the
general principle of relativity.

It should then be also clear that the phenomenon of gravitation is
incorporated in this framework as the concept of transformation is
``applicable'' in all the relevant situations.

This then brings us to the important issue of the appropriate
mathematical formulation for the new theory whose certain
characteristics we have considered above.

\section{Quantum aspects and the Mathematical Foundation for the New Theory}
Now, quantum theory acknowledges an important limitation of the
classical newtonian ideas by recognizing that any observation of a
physical system involves, necessarily, an uncontrollable
disturbance of that physical system. [This is similar in spirit to
special relativity acknowledging the fact that we ordinarily use
electromagnetic radiation to observe material bodies.]

It is then important to recognize that an ``act of observation''
will, necessarily, involve the transformation of the points of the
underlying space representing the physical system that is being
observed. Therefore, transformations of the underlying space are
the key to quantum aspects.

But, we must first realize that ``material bodies'' and the
``physical geometry of space'' aught to be {\em
indistinguishable}.

Firstly, moving a material body from its given ``location'' should
cause changes to the construction of the coordinate system. Motion
of a material body will then change the ``physical geometry''
because the construction of the coordinate system is the basis of
the ``metric function'' of the geometry. Hence, the physical
geometry is determined by material bodies.

In turn, material bodies are also determined by the physical
geometry in that ``given the metric function of the geometry'' we
would know how the totality of {\em all\/} the material bodies are
``located'' relative to each other.

In this context, efforts of Rylov \cite{rylov} at defining
physical geometry as mutual dispositions of the totality of
geometric objects are noteworthy. It is proved in \cite{rylov}
that {\em the distance function determines completely the physical
geometry, and one does not need any additional information for
determination of physical geometry.}

However, the issue remains of the physical characteristics of
material bodies such as, for example, their inertia, electrostatic
charge etc. These are the ``qualities'' of every geometric object
which is to be viewed, necessarily, as a {\em measurable set\/} of
the underlying measure space \cite{srivastava}.

A chosen measure can be averaged over the geometrically
well-defined size of the geometric object under considerations.
This averaged measure provides the non-singular notion of a
point-object with the physical characteristics.

But, the location of the point-object so defined is {\em
indeterminate\/} within the limits of the geometric object.
Consequently, when the measurement of any characteristics such as
the location of this point-object is performed, an {\em intrinsic
indeterminacy\/} is involved. This is expected to lead to various
quantum aspects. See, for further related details, \cite{issues,
field-program, smw-indeterminacy, 100-yrs, heuristic}.

Evidently, these ideas are then fundamentally different from those
of Newton's theory as well as from those of special relativity. In
these new considerations, the concept of ``force'' is abandoned
and is replaced by suitable properties of the geometry - its
transformations. Since ``force'' can be looked upon as a
transformation of the points of the underlying geometry, results
of Newton's theory and those of the special relativity are
obtainable \cite{results} within the proposed formalism.

\acknowledgements I am very much grateful to the referees of the
journal {\em General Relativity \& Gravitation\/} for having
raised various queries while reviewing the submission
\cite{smw-indeterminacy}.

\end{document}